\documentstyle[12pt]{article}

\begin{document}

\topmargin 0pt
\oddsidemargin 5mm

\setcounter{page}{0}
\begin{titlepage}
\vspace{2cm}
\begin{center}

{RADIATION OF RELATIVISTIC CHARGED PARTICLES
IN A SYSTEM WITH ONE DIMENSIONAL RANDOMNESS}

{\large Zh.S.Gevorkian}\\
\vspace{1cm}
{\em Institute of Radiophysics and Electronics}\\
{Ashtarak-2, 378410, Armenia}\\
\vspace{5mm}
{\bf{Abstract}}
\end{center}
Radiation of relativistic charged particles in a system of
randomly spaced plates is considered in the paper. It is shown that
for large number of plates ($N \gg 1$), in the wavelength range
$\lambda \ll l\ll L$ ( where $l$ is the photon mean free path and
$L$ is the system characteristic size) and for angles $|\cos
\theta |\gg (\lambda/2\pi l)^{1/3}$, pseudophoton diffusion
represents the major mechanism of radiation. Total intensity of
radiation is investigated and its strong dependence on the
particle energy and plate number is obtained.
\vfill
\end{titlepage}
\section{Introduction}
\indent
Charged particle radiation in layered media has been considered in many
papers (see e.g. \cite{ML} and \cite{GM} and references therein).
The interest in these systems is caused by the possibility of
their use as high energy particle  detectors
\cite{GM}.Detecting properties of these systems  are based on
the transition radiation.
Transition radiation originating in such systems can be explained
in the following way (see \cite{ML} and \cite{GM}).
A charge moving in a medium creates an  electromagnetic field
(a pseudophoton), which is scattered by  the inhomogeneities of
dielectric permittivity and converted into radiation. The key
problem is account correctly for the scattering of
pseudophotons on the inhomogeneities.

In earlier articles (see, for example , \cite{GM}) which have
addressed to the problem of radiation of relativistic charged
particles in a system of plates embedded in a homogeneous medium
the reflection of the electromagnetic field by an individual plate is
neglected.
However from experience with
three dimensional random media \cite{ZH1} and \cite{ZH} we know that the
multiple scattering of electromagnetic fields plays an essential
role. This role is particularly important in the optical region
which we have mainly in mind.

In the present  paper we consider multiple
scattering effects (taking into account also 
reflection) when a charged particle radiates passing through
a one-dimensional  random medium. Such media can be, in particular, 
those systems in which the plates are randomly spaced in a
homogeneous medium.

It turns out that multiple scattering of the pseudophoton leads
to its diffusion is the dominant in the medium and this diffusion 
contribution to
the radiation intensity. The diffusion contribution
leads to a strong dependence of the radiation intensity on particle energy
and plate number, a fact that is important for the detecting properties 
of the
system. Note that the diffusion contribution is absent in an ordered stack
of plates.

\section{Formulation of the Problem}
\indent

The system considered in the paper consits of a stack of plates
randomly spaced in a homogeneous medium. Let the plates
fill the regions $z_i-a/2 <z<z_i+a/2$ (where $a$ is the plate thickness
and $z_i$ are random coordinates). The dielectric permittivity of the system
may be represented in the following form:
\begin{equation}
\label{BA}
\varepsilon(z,\omega )=\varepsilon  _0(\omega)
+\sum_{i}\left[b(\omega )-\varepsilon
_0(\omega)\right]\left[|\Theta(z-z_i-a/2)-\Theta(z-z_i+a/2)|\right],
\end{equation}
where $\varepsilon  _0(\omega)$ and $b(\omega)$ are respectively
die\-lec\-tric permittivities of the homogeneous me\-di\-um and
of the plates, and $\Theta$ is a step function.
It is convenient to represent the dielectric
permittivity as a sum of average and fluctuating parts:
\begin{equation}
\label{BC}
\varepsilon (z,\omega)=\varepsilon+\varepsilon _r(z,\omega),\quad
\quad < \varepsilon _r(z,\omega)>=0,
\end{equation}
where $\varepsilon =<\varepsilon(z,\omega)>$, $\varepsilon_r \ll \varepsilon$ 
and
averaging over the random coordinates of  plates is determined as
follows
\begin{equation}
\label{BD}
<f(z,\omega)>=\int \prod_{i}\frac{dz_i}{L_z}f(z,{z_i},\omega ),
\end{equation}
where $L_z$ is the system size in the $z$-direction.
In the frequency domain, Maxwell's equations have the following form:
\begin{eqnarray}
\label{BE}
&& {\bf \vec{\nabla}}\times \vec{E}(\vec{r},\omega )=\frac{i\omega
}{c}\vec{B}(\vec{r},\omega ),\quad {\bf \vec{\nabla}}\times \vec{B}(\vec{r},\omega )=
\frac{4\pi e}{c}\,\frac{\vec{v}}{v}\delta  (x)\delta  (y)
e^{i\omega z/v}-\frac{i\omega
}{c}\vec{D}(\vec{r},\omega)\nonumber\\
&& {\bf \vec{\nabla}}\cdot \vec{D}(\vec{r},\omega )=\frac{4\pi e}{v}\delta (x)\delta (y)
e^{i\omega z/v}\quad, {\bf \vec{\nabla}}\cdot \vec{B}(\vec{r},\omega )=0, \quad
\vec{D}(\vec{r},\omega )= \varepsilon (z,\omega)
\vec{E}(\vec{r},\omega)
\end{eqnarray}
Here $\vec{v}\parallel \hat{z}$ is the velocity of the particle.
For convenience we introduce the potentials of electromagnetic field.
\begin{equation}
\label{BF}
\vec{E}(\vec{r},\omega)=\frac{i\omega}{c}\vec{A}(\vec{r},\omega)
-{\rm \nabla} \varphi (\vec{r},\omega)
\end{equation}
Using (\ref{BE}) and (\ref{BF}),  we obtain the equation for $\vec{A}
(\vec{r},\omega)$
\begin{equation}
\label{BG}
\nabla^2\vec{A}+\frac{\omega^2}{c^2}\vec{A}(\vec{r},\omega)
\varepsilon (\vec{r},\omega)-{\rm \nabla}\left[{\bf \vec{\nabla}}\cdot \vec{A}-\frac{i\omega}{c}\varepsilon(\vec{r},\omega)
\varphi (\vec{r},\omega)\right]=\vec{j}(\vec{r},\omega),
\end{equation}
where $\vec{j}(\vec{r},\omega)$ is the Fourier transform of the current of the charged particle
\begin{equation}
\label{BH}
\vec{j}(\vec{r},\omega)=-\frac{4\pi e}{c}\,\frac{\vec{v}}{v}
\delta(x)\delta(y) e^{i\omega z/v}
\end{equation}
\indent
Imposing the Lorentz gauge condition on the potentials, we finally
obtain finally
\begin{equation}
\label{BJ}
{\bf \vec{\nabla}}\cdot \vec{A}-\frac{i\omega}{c}\varepsilon(\vec{r},\omega)
\varphi (\vec{r},\omega)=0; \quad \nabla^2\vec{A}+\frac{\omega^2}{c^2}
\varepsilon (\vec{r},\omega)\vec{A}(\vec{r},\omega)=\vec{j}
(\vec{r},\omega)
\end{equation}
It follows from the symmetry of the problem that the vector
potential $\vec{A}$ is directed along the $z$: 
$A_i=\delta_{\hat{z}i}A(\vec{r},\omega)$.

\section{Radiation Tensor}
\indent
As usual, we decompose the electric field into two parts,
$\vec{E}=\vec{E}_0+\vec{E}_r$. Here $\vec{E}_0$ is the electric
field of the charged particle moving in homogeneous medium with
dielectric permittivity $\varepsilon $, and $E_r$ is the radiation
field caused by fluctuations in the dielectric permittivity. We
define the radiation tensor as follows
\begin{equation}
\label{CA}
I_{ij}(\vec{R})=E_{ri}(\vec{R})E_{rj}^*(\vec{R})
\end{equation}
Here $\vec{R}$ is the radius-vector to the observation point,
which is far from the system, $R\gg L$. The vector
potential is decomposed in a similar way:
$\vec{A}=\vec{A}_0+\vec{A}_r$, where $\vec{A}_0$ and
$\vec{A}_r$, as follows from (ref{BC}) and (\ref{BJ}), satisfy the equations
\begin{eqnarray}
\label{CD}
&&\nabla^2\vec{A}_0+\frac{\omega^2}{c^2}\varepsilon\vec{A}_0=
\vec{j}(\vec{r},\omega)\nonumber \\
&&\nabla^2\vec{A}_r+\frac{\omega^2}{c^2}\varepsilon\vec{A}_r+
\frac{\omega^2}{c^2}\varepsilon_r\vec{A}_r=-\frac{\omega^2}{c^2}
\varepsilon_r\vec{A}_0
\end{eqnarray}
It is convenient to express the radiation  intensity in terms of the
radiation potential $\vec{A}_r$
\begin{eqnarray}
\label{CF}
&&<I_{ij}(\vec{R})>=\frac{\omega^2}{c^2}\delta_{\hat{z}i}\delta
_{\hat{z}j}<A_r(\vec{R},\omega)A_r^*(\vec{R},\omega)>
+\frac{\delta_{\hat{z}i}}{\varepsilon}<A_r(\vec{R},\omega)
\frac{\partial^2}{\partial R_j\partial z}
A_r^*(\vec{R},\omega)>\nonumber \\
&&+\frac{\delta_{\hat{z}j}}{\varepsilon}<A_r^*(\vec{R},\omega)
\frac{\partial^2}{\partial R_i\partial z}
A_r(\vec{R},\omega)>+\frac{c^2}{\omega ^2 \varepsilon^2}<
\frac{\partial^2}{\partial R_i\partial z}
A_r(\vec{R},\omega)\frac{\partial^2}{\partial R_j\partial z}
A_r^*(\vec{R},\omega)>
\end{eqnarray}
In obtaining (\ref{CF}) we assumed that the fluctuations of
dielectric permittivity are much smaller than its mean value
$\varepsilon_r\ll \varepsilon $.

To carry out averaging over the random coordinates of plates, we express
the radiation potential $A_r$ in terms of the Green's function of
the equation (\ref{CD})
\begin{eqnarray}
\label{CE}
&& A_r(\vec{R})=-\frac{\omega ^2}{c^2} \int \varepsilon_r(\vec{r})
A_0(\vec{r})G(\vec{R},\vec{r})d\vec{r}\nonumber\\
&&\left[ \nabla^2+k^2+\frac{\omega ^2}{c^2} \varepsilon_r(z)\right]
G(\vec{r},\vec{r}^\prime)= \delta(\vec{r}-\vec{r}^\prime),
\end{eqnarray}
where $k=\omega\sqrt{\varepsilon}/c$.

\section{Green's  Function}
\indent
The bare Green's  function of equation (\ref{CE}) satisfies the
equation
\begin{equation}
\label{CG}
\left[\nabla^2+k^2+i\delta
\right]G_0(\vec{r}-\vec{r}^\prime)=\delta (\vec{r}-\vec{r}^\prime)
\end{equation}
where $i\delta$, as usual, is an infinitizimal imaginary term.
Solution in the momentum representation has the form:
\begin{equation}
\label{CH}
G_0(\vec{q})=\frac{1}{k^2-q^2+i\delta}
\end{equation}
In the coordinate representation, one has
\begin{equation}
\label{CL}
G_0(r)=-\frac{1}{4\pi  r}e^{ikr}
\end{equation}
To perform the averaging, we use the impurity-diagram method \cite{AA}.
Summing the diagrams in the ladder
approximation, we obtain Dyson's equation for the average Green's  function

\begin{equation}
\label{CM}
\unitlength=1mm
\linethickness{1.5pt}
\begin{picture}(15.00,20.00)
\put(0,5){\vector(1,0){12}}
\put(5,0){\shortstack{$\vec{q}$}}
\put(14,4){\shortstack{$=$}}
\end{picture}
\linethickness{.5pt}
\begin{picture}(60.00,20.00)
\put(2,5){\vector(1,0){12}}
\put(14,5){\vector(1,0){28.3}}
\put(14.2,5){\line(1,2){1.8}}
\put(16.8,9.4){\line(1,1){3.8}}
\put(21.6,14){\line(2,1){4}}
\put(26.4,16){\vector(1,0){4}}
\put(31.3,16){\line(2,-1){4}}
\put(36,13.4){\line(1,-1){3.8}}
\put(40.5,9){\line(1,-2){1.8}}
\put(27,0){\shortstack{$\vec{q}-\vec{p}$}}
\put(5,0){\shortstack{$\vec{q}$}}
\put(24,17){\shortstack{$\vec{p}$}}
\linethickness{1.5pt}
\put(42.3,5){\vector(1,0){12}}
\put(48,0){\shortstack{$\vec{q}$}}
\end{picture}
\end{equation}

The dotted line denotes the correlation function of the
one-dimensional random field
\begin{eqnarray}
\label{CN}
&&---=B(\vec{p})=({2\pi })^2\delta  (\vec{p}_\rho )B(|p_z|)\nonumber
\\
&&B(|z-z^\prime|)=\frac{\omega ^4}{c^4}<\varepsilon
_r(z)\varepsilon_r(z^\prime)>
\end{eqnarray}
where $\vec{p}_\rho $ is the transverse component of $\vec{p}$.
The solution of  equation (\ref{CM}) can be represented in following
form:
\begin{equation}
\label{QA}
G(\vec{q})=\frac{1}{G_0^{-1}(\vec{q})-\int\frac{d\vec{p}}{(2\pi
)^3}B(\vec{p})G_0(\vec{q}-\vec{p}) }
\end{equation}
Using expression (\ref{CH}), we obtain for the averaged Green's
function the following expression:
\begin{equation}
\label{QB}
G(\vec{q})=\frac{1}{k^2-q^2+i{\rm Im}\Sigma(\vec{q})},
\end{equation}
in which the imaginary part ${\rm Im}\Sigma$ of the self-energy is
determined by Ward's identity
\begin{eqnarray}
\label{QC}
&&{\rm Im}\Sigma(\vec{q})=\int \frac {d\vec{p}}{(2\pi
)^3}B(\vec{p}){\rm Im}
G_0(\vec{q}-\vec{p})=\frac{1}{4\sqrt{k^2-q^2_\rho }}\nonumber \\
&&\left[ B(|q_z-\sqrt{k^2-q^2_\rho }|)+B (|q_z+\sqrt
{k^2-q^2_\rho }|)\right],\quad |\vec{q}_\rho |<k
\end{eqnarray}
The decay length of  pseudophoton in the $z$ direction
 is determined by the
imaginary part  of the Green's function, in the following way (see.,
e.g. \cite{VA})
\begin{equation}
\label{QD}
l(\vec{q})=\frac{\sqrt{k^2-q^2_\rho }}{{\rm Im}\Sigma(\vec{q})}
\end{equation}
As one could expect, the decay length depends on the pseudophoton
momentum direction. In the case where the momentum is
directed along $z$, one obtains from (\ref{QD}) and (\ref{QC})
\begin{equation}
\label{QE}
l(\theta=0 )=\frac{4k^2}{B(0)+B(2k)}
\end{equation}
We shall call this quantity the pseudophoton mean free path.

Using (\ref{BA}), (\ref{BC}) and (\ref{CN}) one finds for
correlation function
\begin{equation}
\label{QK}
B(q_z)=\frac{4(b-\varepsilon)^2 n \sin^2q_z
a/2}{q_z^2}\frac{\omega ^4}{c^4}
\end{equation}
Here $n=N/L_z$ is concentration of plates in the system. From
(\ref{QK}) it
follows  that $B(0)=\omega ^4/c^4\times(b-\varepsilon)^2
na^2 $. On the other hand, when $ka\gg 1$
, $B(2k)/B(0)\sim 1/(ka)^2 \ll 1$. Therefore the
photon mean free path is
\begin{equation}
\label{QV}
l\equiv l(\theta =0)\approx \left\{ \begin {array}{ll}
4k^2/B(0),&\,ka\gg 1\\
2k^2/B(0), &\, ka\ll 1 \end{array} \right.
\end{equation}
The calculation carried out above is correct only in the weak
scattering regime, when
\mbox{$\displaystyle\frac{{\rm
Im}\Sigma(\vec{q})}{k^2-q^2_\rho}\ll 1$.}
Using (\ref{QC}) we obtain
\begin{equation}
\label{QN}
\frac{B(0)+B(2k|\cos\theta|)}{4k^3 |\cos\theta|^3}\ll 1
\end{equation}
From (\ref{QN}) it follows that at $\theta \approx\pi /2$ the
condition of weak scattering is not satisfied. This is natural,
because in this case the pseudophoton moves parallel to the
plates. Taking $\theta =\pi /2-\delta$ and using
(\ref{QE}) and (\ref{QN}), one has $\delta  \gg (1/kl)^{1/3}$.

\section{Radiation Intensity in the Single Scattering
Approximation}
\indent

Substitution of (\ref{CE}) into (\ref{CF}) gives the following
expression for radiation tensor
\begin{eqnarray}
\label{QM}
&&I_{ij} (\vec{R})=\delta  _{\hat{z}i}\delta _{\hat{z}j}
\frac{\omega ^6}{c^6}\int d\vec{r}d\vec{r}{\,}^\prime
A_0(\vec{r})A_0^*(\vec{r}{\,}^\prime)<\varepsilon_r(\vec{r})
\varepsilon
_r(\vec{r}{\,}^\prime)G(\vec{R},\vec{r})G^*(\vec{r}{\,}^\prime,\vec{R})>
\nonumber \\
&&+\frac{\omega ^2}{c^2}\,\frac{1}{\varepsilon  ^2}
\int d\vec{r}d\vec{r}{\,}^\prime
A_0(\vec{r})A_0^*(\vec{r}{\,}^\prime)<\varepsilon  _r(\vec{r})
\varepsilon
_r(\vec{r}{\,}^\prime)\frac{\partial^2}{\partial R_i\partial z}
G(\vec{R},\vec{r})\frac{\partial^2}{\partial R_j\partial z}
G^*(\vec{r}{\,}^\prime,\vec{R})>\nonumber \\
&&+\delta  _{\hat{z}j}\frac{\omega ^4}{c^4\varepsilon}
\int d\vec{r}d\vec{r}{\,}^\prime A_0(\vec{r})A_0^*(\vec{r}{\,}^\prime)
<\varepsilon  _r(\vec{r})
\varepsilon_r(\vec{r}{\,}^\prime)G^*(\vec{r},\vec{R})
\frac{\partial^2}{\partial R_i\partial z}
G(R,r)>\nonumber \\
&&+\delta  _{\hat{z}i}\frac{\omega ^4}{c^4\varepsilon}
\int d\vec{r}d\vec{r}{\,}^\prime A_0(\vec{r})A_0^*(\vec{r}{\,}^\prime)
<\varepsilon  _r(\vec{r})
\varepsilon_r(\vec{r}{\,}^\prime)G(\vec{R},\vec{r})
\frac{\partial^2}{\partial R_j\partial z}G^*(\vec{r},\vec{R})>
\end{eqnarray}

In the single scattering approximation, we substitute the Green's functions
appearing in (\ref{QM}) by bare one (\ref{CL}) functions. Since the
observation point $\vec{R}$ is far from radiating
system, one finds using (\ref{CL}) the following useful relations
\begin{equation}
\label{PA}
G_0(\vec{R},\vec{r})\approx-\frac{1}{4\pi
R}e^{ik(R-\vec{n}\vec{r})},\,\frac{\partial^2 G_0(\vec{R},\vec{r})}
{\partial R_i\partial z}\approx \frac{k^2n_in_z}{4\pi R}
e^{ik(R-\vec{n}\vec{r})},\,R\gg r
\end{equation}
Here $\vec{n}$ is the unit vector in the direction
of observation point $\vec{R}$. Inserting (\ref{PA}) into
 (\ref{QM}) and using
(\ref{CN}), for the radiation tensor we find
\begin{eqnarray}
\label{PB}
&& I^0_{ij}(\vec{R})=\frac{\omega ^2}{c^2}\,\frac{1}{16\pi ^2R^2}
\int d\vec{r}d\vec{r}{\,}^\prime e^{ik\vec{n}(\vec{r}-\vec{r}{\,}^\prime)}
B(|z-z^\prime|) A_0(\vec{r})A_0^*(\vec{r}{\,}^\prime)\nonumber \\
&&\left[ \delta _{\hat{z}i}\delta  _{\hat{z}j}-\delta
_{\hat{z}i} n_j n_z-\delta  _{\hat{z}j}n_i n_z+n_in_jn^2_z\right]
\end{eqnarray}
By solving (\ref{CD}), we easily obtain
\begin{equation}
\label{PC}
A_0(\vec{q})=-\frac{8\pi^2 e}{c}\,\frac{\delta  (q_z-\omega
/v)}{k^2-q^2}
\end{equation}
Using (\ref{PC}) in (\ref{PB}) and integrating,
we find the radiation
intensity $I(\vec{n})=\frac{c}{2}R^2 I_{ii}(\vec{R})$ in the
single scattering approximation:
\begin{equation}
\label{PD}
I^0(\vec{n})=\frac{\pi e^2}{c}\,{\delta
(0)}\frac{B(|k_0-kn_z|)n_\rho
^2}{\left[k^2n^2_z-k_0^2\right]^2}\,\frac{\omega ^2}{c^2}
\end{equation}
Here $k_0=\omega /v$ while the $\delta  $-type singularity of
(\ref{PD}) is caused by the infinite path of the charged
particle in the medium. If one takes into account the finite size
of the system, $\delta(0)$ must be replaced by
$L_z/2\pi$.
To analyse the angular dependence of (\ref{PD}), it is
convenient to represent it in the form
\begin{equation}
\label{PF}
I^0(\theta )=\frac{e^2}{2c}\frac{L_z B(|k_0-k\cos\theta
|)\sin^2\theta }{{\left[\gamma ^{-2}+\sin^2\theta\,k^2/k^2_0
\right]}^2}\,\frac{\omega ^2}{k_0^4 c^2}
\end{equation}
Here $\gamma ={(1-\varepsilon v^2/c^2)}^{-1/2}$ is the Lorentz
factor of the particle. Note some features of the expression (\ref{PF}):
For relativistic energies ($\gamma \gg 1, k_0\rightarrow k$), the
radiation intensity in the forward direction, for short waves
$ka\gg 1$, is significantly
higher than in the backward direction. The maximum
lies in the range of angles $\theta \sim \gamma ^{-1}$. This result
is consistent with the results of \cite{ML} and \cite{GM}. Since
$B\sim n$, the radiation intensity in this approximation, as
one should expect, is proportional to the total number $N$ of
plates in the system.

\section{Diffusion Contribution to the Radiation Intensity}
\indent

In the diffusion approximation, the averages appearing in
(\ref{QM}) are determined by the following diagrams
\begin{eqnarray}
\label{PP}
&&<G(\vec{R},\vec{r})G^*(\vec{r}{\,}^\prime,\vec{R})>^D=
\raisebox{-1.7cm}{\setlength{\unitlength}{.3cm}
\begin{picture}(15,11)
\put(11,4){\vector(-1,0){10}}
\put(1,8){\vector(1,0){10}}
\multiput(5,4)(0,0.5){7}{\line(3,1){3}}
\multiput(5,4)(3,0){2}{\line(0,1){4}}
\put(6.5,4){\line(3,1){1.5}}
\put(5,7.5){\line(3,1){1.5}}
\put(1,2.5){\shortstack{$\vec{R}$}}
\put(1,8.5){\shortstack{$\vec{R}$}}
\put(11,2.5){\shortstack{$\vec{r'}$}}
\put(11,8.5){\shortstack{$\vec{r}$}}
\put(5,2.5){\shortstack{$\vec{r}_2$}}
\put(8,8.5){\shortstack{$\vec{r}_3$}}
\put(5,8.5){\shortstack{$\vec{r}_1$}}
\put(8,2.5){\shortstack{$\vec{r}_4$}}
\end{picture}}\nonumber \\
&&<G(\vec{R},\vec{r})\frac{\partial ^2}{\partial R_j\partial z}
G^*(\vec{r}{\,}^\prime,\vec{R})>^D=
\raisebox{-1.7cm}{\setlength{\unitlength}{.3cm}
\begin{picture}(15,11)
\put(11,4){\vector(-1,0){10}}
\put(1,8){\vector(1,0){10}}
\multiput(5,4)(0,0.5){7}{\line(3,1){3}}
\multiput(5,4)(3,0){2}{\line(0,1){4}}
\put(6.5,4){\line(3,1){1.5}}
\put(5,7.5){\line(3,1){1.5}}
\put(0,2.5){\shortstack{$
\frac{\partial ^2}{\partial R_j\partial z}$}}
\put(1,4.5){\shortstack{$\vec{R}$}}
\put(1,8.5){\shortstack{$\vec{R}$}}
\put(11,2.5){\shortstack{$\vec{r'}$}}
\put(11,8.5){\shortstack{$\vec{r}$}}
\put(5,2.5){\shortstack{$\vec{r}_2$}}
\put(8,8.5){\shortstack{$\vec{r}_3$}}
\put(5,8.5){\shortstack{$\vec{r}_1$}}
\put(8,2.5){\shortstack{$\vec{r}_4$}}
\end{picture}} \nonumber \\
&&<\frac{\partial ^2}{\partial R_i\partial z}G(\vec{R},\vec{r})
G^*(\vec{r}{\,}^\prime,\vec{R})>^D=
\raisebox{-1.7cm}{\setlength{\unitlength}{.3cm}
\begin{picture}(15,11)
\put(11,4){\vector(-1,0){10}}
\put(1,8){\vector(1,0){10}}
\multiput(5,4)(0,0.5){7}{\line(3,1){3}}
\multiput(5,4)(3,0){2}{\line(0,1){4}}
\put(6.5,4){\line(3,1){1.5}}
\put(5,7.5){\line(3,1){1.5}}
\put(0,9){\shortstack{$
\frac{\partial ^2}{\partial R_i\partial z}$}}
\put(1,2.5){\shortstack{$\vec{R}$}}
\put(1,6.5){\shortstack{$\vec{R}$}}
\put(11,2.5){\shortstack{$\vec{r'}$}}
\put(11,8.5){\shortstack{$\vec{r}$}}
\put(5,2.5){\shortstack{$\vec{r}_2$}}
\put(8,8.5){\shortstack{$\vec{r}_3$}}
\put(5,8.5){\shortstack{$\vec{r}_1$}}
\put(8,2.5){\shortstack{$\vec{r}_4$}}
\end{picture}} \nonumber \\
&&<\frac{\partial ^2}{\partial R_i\partial z}G(\vec{R},\vec{r})
\frac{\partial ^2}{\partial R_j\partial z}
G^*(\vec{r}{\,}^\prime,\vec{R})>^D=
\raisebox{-1.7cm}{\setlength{\unitlength}{.3cm}
\begin{picture}(15,11)
\put(11,4){\vector(-1,0){10}}
\put(1,8){\vector(1,0){10}}
\multiput(5,4)(0,0.5){7}{\line(3,1){3}}
\multiput(5,4)(3,0){2}{\line(0,1){4}}
\put(6.5,4){\line(3,1){1.5}}
\put(5,7.5){\line(3,1){1.5}}
\put(0,2.5){\shortstack{$
\frac{\partial ^2}{\partial R_j\partial z}$}}
\put(0,9){\shortstack{$
\frac{\partial ^2}{\partial R_i\partial z}$}}
\put(1,4.5){\shortstack{$\vec{R}$}}
\put(1,6.5){\shortstack{$\vec{R}$}}
\put(11,2.5){\shortstack{$\vec{r'}$}}
\put(11,8.5){\shortstack{$\vec{r}$}}
\put(5,2.5){\shortstack{$\vec{r}_2$}}
\put(8,8.5){\shortstack{$\vec{r}_3$}}
\put(5,8.5){\shortstack{$\vec{r}_1$}}
\put(8,2.5){\shortstack{$\vec{r}_4$}}
\end{picture}}
\end{eqnarray}
Here the filled rectangle corresponds to the diffusion propagator
\begin{equation}
\label{GG}
P(\vec{r}_1,\vec{r}_2,\vec{r}_3,\vec{r}_4)=
\raisebox{-1.7cm}{\setlength{\unitlength}{.3cm}
\begin{picture}(5,11)
\put(4,4){\vector(-1,0){3}}
\put(1,8){\vector(1,0){3}}
\multiput(1,4)(0,0.5){7}{\line(3,1){3}}
\multiput(1,4)(3,0){2}{\line(0,1){4}}
\put(2.5,4){\line(3,1){1.5}}
\put(1,7.5){\line(3,1){1.5}}
\put(0.5,2.5){\shortstack{$\vec{r}_2$}}
\put(3.5,8.5){\shortstack{$\vec{r}_3$}}
\put(0.5,8.5){\shortstack{$\vec{r}_1$}}
\put(3.5,2.5){\shortstack{$\vec{r}_4$}}
\end{picture}}
=\sum
\raisebox{-1.7cm}{
\setlength{\unitlength}{.3cm}
\begin{picture}(15,10)
\put(11,4){\vector(-1,0){10}}
\put(1,8){\vector(1,0){10}}
\multiput(1,4)(0,0.5){8}{\line(0,1){0.4}}
\multiput(3,4)(0,0.5){8}{\line(0,1){0.4}}
\multiput(5,4)(0,0.5){8}{\line(0,1){0.4}}
\multiput(7,4)(0,0.5){8}{\line(0,1){0.4}}
\multiput(9,4)(0,0.5){8}{\line(0,1){0.4}}
\multiput(11,4)(0,0.5){8}{\line(0,1){0.4}}
\put(1,2.5){\shortstack{$\vec{r}_2$}}
\put(1,8.5){\shortstack{$\vec{r}_1$}}
\put(11,2.5){\shortstack{$\vec{r}_4$}}
\put(11,8.5){\shortstack{$\vec{r}_3$}}
\end{picture}}
\end{equation}
Using (\ref{QM}),(\ref{PP}) and (\ref{GG}), we obtain the
following expression for the diffusion contribution
\begin{eqnarray}
\label{PB1}
&& I^D_{ij}(\vec{R})=\frac{k ^2}{16\pi^2 R^2\varepsilon}\,
\int d\vec{r}d\vec{r}{\,}^\prime
B(r-r^\prime) A_0(\vec{r})A_0^*(\vec{r}{\,}^\prime)\,\int
d\vec{r}_1d\vec{r}_2 d\vec{r}_3 d\vec{r}_4\nonumber \\
&&e^{-ik\vec{n}(\vec{r}_1-\vec{r}_2)}P(\vec{r}_1,\vec{r}_2,
\vec{r}_3,\vec{r}_4)G(\vec{r}_3,\vec{r})G^*(\vec{r},
\vec{r}_4)\nonumber\\
&&\left[ \delta  _{\hat{z}i}\delta  _{\hat{z}j}+
n_i n_j n_z^2-\delta _{\hat{z}i}n_jn_z-
\delta_{\hat{z}j}n_i n_z\right]
\end{eqnarray}
The diffusion propagator $P$ which appears in (\ref{PB1}) is found
similarly to the three dimensional case \cite{ZH}. It follows
from (\ref{GG}) that $P(\vec{r}_1,\vec{r}_2,\vec{r}_3,\vec{r}_4)$
can be represented in form
\begin{equation}
\label{PX}
P(\vec{r}_1,\vec{r}_2,\vec{r}_3,\vec{r}_4)=B(\vec{r}_1-
\vec{r}_2)B(\vec{r}_3-\vec{r}_4)P(\vec{R}^\prime,\vec{r}_1
-\vec{r}_2,\vec{r}_3-\vec{r}_4)
\end{equation}
where $\vec{R}^\prime=\frac{1}{2}(\vec{r}_3+\vec{r}_4-\vec{r}_1-\vec{r}_2)$
and $P$ satisfies the equation
\begin{equation}
\label{PL}
\int\frac{d\vec{p}}{(2\pi )^3}\left[ 1-\int\frac{d\vec{q}}
{(2\pi )^3}f(\vec{q},\vec{K})B(\vec{p}-\vec{q})\right]
P(\vec{K},\vec{p},\vec{q}{\,}^\prime)=f(\vec{q}{\,}^\prime,\vec{K})
\end{equation}
Here
\begin{equation}
\label{PM}
f(\vec{q},\vec{K})=G(\vec{q}+\vec{K}/2)G^*(\vec{q}-\vec{K}/2)
\end{equation}
As it will be seen further, one has to know
$P$ when $\vec{K}\rightarrow 0$. In this limit, the diffusion
propagator has the form \cite{ZH}
\begin{equation}
\label{PN}
P(\vec{K}\rightarrow 0,\vec{p},\vec{q})=\frac{{\rm Im}G(\vec{p})
{\rm Im}G(\vec{q})}{{\rm Im}\Sigma(\vec{q})}A(\vec{K})
\end{equation}
where
\begin{equation}
\label{FA}
A(\vec{K})=\left[ 3\int\frac{(\vec{q}\vec{K})^2 {\rm Im}G(\vec{q})}
{{\rm Im}^2\Sigma(\vec{q})}\,\frac{d\vec{q}}{(2\pi )^3}\right]^{-1}
\end{equation}
Choosing $\vec{K}\parallel\hat{z}$ and using (\ref{QC}), we obtain
\begin{equation}
\label{FB}
A(K)=\left[\frac{6K^2k^5}{\pi }\int\limits_{-1}^{1}\frac{dx
x^4}{\left[B(0)+B(2k|x|) \right]^2} \right]^{-1}
\end{equation}
Here we have changed
variables while integrating over the angles. It follows from the form
of correlation function (\ref{QK}) that the main
contribution into the integral(\ref{FB}) is given by the values of
$x$ close to unity (the corresponding angles are close to zero).
Taking into account this fact, for $A(K)$, we have approximately
\begin{equation}
\label{FC}
A(K)=\frac{1}{k}\,\frac{20\pi}{3K^2 l^2},
\end{equation}
where $l=4k^2/B(0)$ is the pseudophoton's mean free path. In the
expression for radiation intensity it is convenient to turn to
new variables of integration
\begin{equation}
\label{FD}
 \vec{R}'=\frac{1}{2}(\vec{r}_3+\vec{r}_4-\vec{r}_1-\vec{r}_2),\quad
\vec{x}_1=\vec{r}_1-\vec{r}_2,\quad
\vec{x}_2=\vec{r}_3-\vec{r}_4,
\quad \vec{r}_4\equiv\vec{r}_4,
\end{equation}
which gives
\begin{equation}
\label{FE}
I^D_{ij}(\vec{R})=\frac{k ^2}{16\pi^2 R^2\varepsilon}
\left(\delta  _{\hat{z}i}\delta  _{\hat{z}j}+
n_i n_j n_z^2-\delta _{\hat{z}i}n_j n_z-
\delta_{\hat{z}j}n_i n_z\right)D,
\end{equation}
where $D$ is given by the expression
\begin{eqnarray}
\label{FG}
&&D=\int d\vec{r}d\vec{r}{\,}^\prime d\vec{R}'d\vec{x}_1 d\vec{x}_2
d\vec{r}_4
A_0(\vec{r})B(r-r^\prime) A_0^*(\vec{r}{\,}^\prime)
e^{-ik\vec{n}\vec{x}_1}\nonumber \\
&&B(x_1)B(x_2)
P(\vec{R}',\vec{x}_1,\vec{x}_2)
G(\vec{x}_2+\vec{r}_4-\vec{r})G^*(\vec{r}{\,}^\prime-\vec{r}_4)
\end{eqnarray}
In the Fourier representation (\ref{FG}) has the following form
\begin{eqnarray}
\label{FF}
&&D=\int \frac{d\vec{q}_1d\vec{q}_2 d\vec{q}_3d\vec{q}_4}
{(2\pi )^{12}}|A_0(\vec{q}_1)|^2 B(\vec{q}_2)B(\vec{q}_3)
B(\vec{q}_4)\nonumber \\
&&P(K'\rightarrow 0,
-\vec{q}_3-k\vec{n},\vec{q}_1+\vec{q}_2+ \vec{q}_4)
|G(\vec{q}_1+\vec{q}_2)|^2
\end{eqnarray}
Substituting (\ref{PN}) into (\ref{FF}) and integrating
(using the Ward identity (\ref{QC})), we shall obtain
\begin{equation}
\label{FL}
D=A(K){\rm Im} \Sigma(k\vec{n})\int\frac{d\vec{q}}{(2\pi )^3}
{\rm } \frac{B(\mid{\sqrt{k^2-q_{\rho}^2}-q_z}\mid)+B(\mid{\sqrt{k^2-q_{\rho}^2}+q_z}\mid)}{B(0)+B(2\sqrt{k^2-q_{\rho}^2})}|A_0(\vec{q})|^2
\end{equation}
Finally, we evaluate the integral over the momentum
remaining in (\ref{FL}). Using (\ref{QC}) and (\ref{PC})
in (\ref{FL}), we have
\begin{eqnarray}
\label{FM}
D&=&A(\vec{K}){\rm Im} \Sigma(k\vec{n})\frac{16\pi ^2
e^2}{c^2}L_z\times\nonumber \\
&\times&\int\frac{d\vec{q}_\rho }{(2\pi )^2}
\frac{1}{(k^2-k_0^2-q^2_{\rho})^2}
\frac{\left[ B\left( |k_0+\sqrt{k^2-q_\rho ^2}|\right)+
B\left( |k_0-\sqrt{k^2-q_\rho ^2}|\right)\right]}
{B(0)+B(2\sqrt{k^2-q_{\rho}^2)}}
\end{eqnarray}
It follows from (\ref{FM}), that for relativistic energies
$k_0\rightarrow k$, the main contribution to
the integral (\ref{FM}) is given by the values $q_\rho \rightarrow 0$.
Taking into account this fact, and the fact that when $\gamma^2
\gg ak$ function $B$ varies slowly as well as (\ref{FC}), we
find
\begin{equation}
\label{FX}
D\approx\frac{e^2}{c^2}\frac{20\pi
^2}{3K^2l^2}L_z\frac{B(0)+B(2k|n_z|)}{k^2}\,\frac{1}{|n_z|}\,\frac{\gamma
^2}{k_0^2}
\end{equation}
Substituting (\ref{FX}) into (\ref{FE}) for the diffusion
contribution into the radiation intensity, we obtain finally
\begin{equation}
\label{FH}
I^D(n_z)=\frac{5}{6}\,\frac{e^2\gamma
^2}{\varepsilon c}\,\left( \frac{L_z}{l(\omega)}\right)^3\,\frac{1-n_z^2}{|n_z|}
\end{equation}
In deriving (\ref{FH}) we substitute $1/K^2\,
{\rm at}\,K\rightarrow 0$ by $L_z^2$ as usual (and also assume that
$L_z\ll L_x, L_y$).
Note some pecularities of the diffusion
contribution. It is easy to verify that $I^D/I^0\sim
L_z^2/l^2\gg 1.$ This means that for $k|\cos\theta |^3l\gg 1$ and
$l\ll L_z$ the diffusion contribution is the major one. As one should
expect, the backward and forward intensities are equal to each
other. Note that with an
accuracy of unimportant numerical coefficients the formula
(\ref{FH}) is correct both for short $ka\gg 1$ and for long $ka\ll
1$ waves. All information on randomness is contained in the mean
free path $l(\omega )$. In the next section we shall specify the
form of $l(\omega )$  in particular cases.

\section{Pseudophoton Mean Free Path}
\indent
The pseudophoton mean free path in our theory is described by the
expression (\ref{QV}). In the impurity diagram method
\cite{AA}, as usual, we don't take into account the diagrams
which correspond to the situation of three or more plates at the
same point. This is valid provided that $|\sqrt{b/\varepsilon
}-1|ka \ll 1$ which means that for scattering of a photon on a
plate, the
Born approximation is fulfilled. However it is well known
\cite{AA} that the formulae are also correct in the general case
provided that one employs the exact scattering amplitude instead
of Born scattering amplitude. In our case this
means that formula (\ref{FH}) is correct  in the general case
provided that a suitable expression is used for pseudophoton mean
free path $l(\omega )$.

The photon mean free path in the medium is related to the
photon transmission coefficient through a plate.
\begin{equation}
\label{RA}
l(\omega)=\frac{\left[ 1-{\rm Re}t(\omega )\right]^{-1}}{n}
\end{equation}
where $t(\omega )$ is the photon transmission coefficient
through a dielectric plate with photon momentum normal to the
plate \cite{VA}
\begin{equation}
\label{RB}
t(\omega )=\frac{2i\sqrt{\frac{b(\omega)}{\varepsilon (\omega )}}
{\rm exp}(-ika)}{\left[ \frac{b(\omega)}{\varepsilon (\omega)
}+1\right]\sin \sqrt{\frac{b(\omega )}{\varepsilon (\omega
)}}ka+2i\sqrt{\frac{b(\omega)}{\varepsilon (\omega )}}\cos
\sqrt{\frac{b(\omega )}{\varepsilon (\omega )}}ka}
\end{equation}
It follows from (\ref{FH}) and (\ref{RA}) that the maximum of
spectral radiation intensity lies in the  frequency region where
transmission coefficient is minimal. It follows from (\ref{RA})
that the minimal value of $l(\omega )$ is $1/n$. Now we shall clarify
the conditions under which this value is achieved. In the Born approximation
$|\sqrt{b/\varepsilon}-1|ka\ll 1$, using (\ref{RB}) and
(\ref{RA}), we obtain
\begin{equation}
\label{RD}
l(\omega
)=\frac{2}{n\left(\sqrt{\frac{b}{\varepsilon}}-1\right)^2k^2a^2}
\end{equation}
which  agrees with (\ref{QV}). More interesting for us is the
geometrical optics region $|\sqrt{\frac{b}{\varepsilon}}-1|ka\gg 1$.
Substituting (\ref{RB}) into (\ref{RA}) and
neglecting the strongly oscillating terms, we have
$l(\omega )\sim 1/n$. Thus in the geometrical optics
region the photon mean free path does not depend on
the frequency, and radiation intensity is maximal.
 Integrating the spectral intensity over angles and frequencies in
 this  region we find that the total intensity depends on the
 particle energy as  $I^t\sim \gamma ^2$.
By contrast the energy dependence of the radiation intensity in typical
transition radiation from a single  interface in the optical region is
logarithmic (see, for example,\cite{GM}). In order to find the
dependence of the radiation intensity on the number of plates, note that
$L_z=N/n$ and from (\ref{FH}) one has $I^t\sim N^3$. One of the
important conditions for the applicability of the theory is the
condition $l\ll L_z$. Substituting $L_z=N/n$ and $l=1/n$ into
this condition we find a condition for plate number $N\gg 1$.

Note that we didn't take into account the
absorption of photons. This is valid provided that
$l\ll \ l_{in}$ (where $l_{in}$ is the photon inelastic mean
free path in the medium). In the theory of diffusive propagation 
the weak absorption $(l \ll l_{in})$ is taken into
account in the following way (see, for example, \cite{PW}). If
the absorption is so weak that $L_z<(l l_{in})^{1/2}$, then 
expression  (\ref{FH}) remains unchanged. When $L_z>(l
l_{in})^{1/2}$ one must substitute $L_z^2$ by $l l_{in}$ in
(\ref{FH})
\begin{equation}
\label{FH1}
I^D(n_z,\omega )=\frac{5}{6}\,\frac{e^2\gamma
^2}{\varepsilon c}\, \frac{L_z l_{in}(\omega)}{l^2(\omega )}\,
\frac{1-n_z^2}{|n_z|}
\end{equation}
It follows from (\ref{FH1})  that in this case the dependence of
radiation intensity on plate number is weaker $I\sim N$.

\section{Conclusions}
\indent
We have considered the diffusion contribution for radiation
intensity of a relativistic particle passing through a stack of
randomly spaced plates. It was shown that for a large number of plates
$N\gg 1$, in the  wavelength region $\lambda \ll l$ and for the
angles $|\cos\theta |\gg (1/kl)^{1/3}$, the diffusion contribution
is the dominant one. Note that the backward and forward intensities of
relativistic charged particle radiation intensity are equal,  whereas
in the regular stack case relativistic particle radiates mainly in the 
forward direction.

Now let us discuss the possible experimental realizations of our
theory. For applicability of the theory the fulfilment of the
following inequalities is necessary $\lambda \ll l(\lambda )\ll
l_{in},L_z$.

The transition radiation of relativistic charged
particles in a stack of plates has been investigated experimentally in many
papers (see, for example, \cite{ID}). Unfortunately in these
papers only the X-ray region was studied. In the X-ray region
 the above-mentioned inequalities are not satisfied. Optical
 transition radiation of relativistic particles has been investigated
 in experimental work \cite{LW}. However in this experiment
 only one or two parallel plates were used.
 Samples in \cite{LW} were prepared by vacuum deposition of
 various metallic coatings (Al, Ag, Au, Cu) on mylar foils
 $3.5\mu m$ thick. Note that these samples are optimal for our
 goals. They ensure minimal transmission due to metallic coatings
 and weakness of absorption due to mylar foils. So it will be
 interesting to investigate experimentally the optical $(\lambda
 \sim2000A^0-6000A^0)$ transition radiation of relativistic
 $(\gamma \sim10^2-10^3 )$ electrons passing through a stack of
 such samples randomly spaced in the vacuum.

I thank V.Arakelyan and Referees for useful comments.
The research described in this paper was made possible in part
by Grant $\#RY2000$ from the International Science Foundation.

\end{document}